\documentclass[nofootinbib,twocolumn,preprintnumbers,amssymb,amsmath,superscriptaddress,letterpaper]{revtex4-1}[12pt]

\usepackage{graphicx}
\usepackage{dcolumn}
\usepackage{bm}
\usepackage{natbib}
\usepackage{pstricks}
\usepackage{epsfig}
\usepackage{epstopdf}
\usepackage{epigraph}
\usepackage{mathtools}
\usepackage[unicode=true,
 bookmarks=false,
 breaklinks=false,pdfborder={0 0 1},backref=false,colorlinks=true,citecolor=blue,urlcolor=blue,linkcolor=blue]
 {hyperref}

\newcommand{\be}{\begin{equation}}
\newcommand{\ee}{\end{equation}}
\newcommand{\bea}{\begin{eqnarray}}
\newcommand{\eea}{\end{eqnarray}}

\begin{document}




\title{Ensemble Average Theory of Gravity}

\author{Nima Khosravi}
\email{n-khosravi@sbu.ac.ir}
\affiliation{Department of Physics, Shahid Beheshti University, G.C., Evin, Tehran 19839, Iran}

\date{\today}

\begin{abstract}
We put forward the idea that all the theoretically consistent models of gravity have contributions to the observed gravity interaction. In this formulation, each model comes with its own Euclidean path-integral weight where general relativity (GR) has automatically the maximum weight in high-curvature regions. We employ this idea in the framework of Lovelock models and show that in four dimensions the result is a specific form of the $f(R,G)$ model. This specific $f(R,G)$ satisfies the stability conditions and possesses self-accelerating solutions. Our model is consistent with the local tests of gravity since its behavior is the same as in GR for the high-curvature regime. In the low-curvature regime the gravitational force is weaker than in GR, which can be interpreted as the existence of a repulsive fifth force for very large scales. Interestingly, there is an intermediate-curvature regime where the gravitational force is stronger in our model compared to GR. The different behavior of our model in comparison with GR in both low- and intermediate-curvature regimes makes it observationally distinguishable from $\Lambda$CDM.

\end{abstract}

\maketitle

\begin{quote}
``What really interests me is whether God had any choice in the creation of the world.''

---\textit{Albert Einstein} 
\end{quote}


{\it{Introduction}}:
General relativity (GR) is now more than one hundred years old but is still mysterious. GR with a cosmological constant (CC) and cold dark matter (CDM), hence $\Lambda$CDM, can model almost all we know about the Universe from the observational data. Although $\Lambda$CDM fits perfectly with the data, it raises a few questions which can be important depending on the
viewpoint and the sensitivity. One question is the direct detection of dark matter (CDM), which is currently a very active field of research in particle and astroparticle physics but is not the question of this work. The nature of dark energy can also be questioned in many ways. The most important technical question about dark energy is the cosmological constant problem, i.e. the fact that highly fine-tuned parameters are required in order to address the observed value of the CC in comparison with the theoretical prediction \cite{CCproblem}. There are different approaches to address the CC problem including the anthropic principle. In a different approach, the solution to the CC problem is tied to the eventual understanding of quantum gravity \cite{qgcc}. Modifying GR is also a way to attack this problem, though it is not the only motivation for modified gravity \cite{mg}. There have been a few claims of discrepancies between the large-scale and the early-Universe data (see, e.g., Ref.~\cite{riess}), which may disfavor $\Lambda$CDM. In addition, Ref.~\cite{Planck-MG} showed that the (phenomenologically formulated) modified gravity models may be favored by the Planck data. We should emphasize that in spite of all the efforts in modifying gravity there has so far been no success in solving the CC problem, and all the models are disfavored by data in comparison to $\Lambda$CDM.\footnote{As far as we know, there only modified gravity model which is as compatible with the data as $\Lambda$CDM and possesses the same number of free parameters, while being distinguishable from $\Lambda$CDM, is the nonlocal model of gravity proposed in Ref.~\cite{non-local}.} Regardless of the frustrating inability of modified gravity to describe the real Universe, there is an interesting and important byproduct of having many different theories and models of gravity at hand. By studying different models we can understand GR better, although there is still a remaining question: why GR should govern the gravitational interactions if there are other theoretically consistent models of gravity? To shed some light on this question we introduce the idea of {\it \"ubermodeling},\footnote{The prefix ``\"uber" is used to emphasize that we work in an space of models, a concept that is a branch of mathematics, named meta-mathematics. Meta-mathematics is about models and not what is inside each model. Since the prefix ``meta" has a definite meaning in mathematics we prefer to not name our model meta-modeling.} as follows.

{\it{\"Ubermodeling}}:
Suppose $\mathbb{M}$ is the space of all the theoretically possible models for a specific system. Within $\mathbb{M}$, one may find models
which are not theoretically consistent, e.g. those that suffer from ghosts or other kinds of instability. Accordingly, we define the space $\mathbb{H}$ which includes all the healthy models of $\mathbb{M}$. Now we put forward the idea of \"ubermodeling as follows: the model which is realized in nature is an ensemble average of all the models in $\mathbb{H}$. This idea is motivated by Wilsonian quantum field theory as well as string theory. It can also be illustrated in the statistical mechanics framework by assuming each model as a micro-state of the system. Then \"ubermodeling says that nature is governed by the thermodynamical properties of such a system, meaning that one needs to take the ensemble average of all the models to reach the true model of nature. From an alternative perspective, the \"ubermodeling idea can be seen in the framework of path integral with properties deeply similar to statistical mechanics. Here we take the Euclidean path integral approach and associate to each model (i.e. path/micro-state) a corresponding probability $e^{-{\cal S}_i}$, where ${\cal S}_i$ is the action corresponding to the $i$'th model. In a very abstract language the model of nature $\cal M$ is given by
\begin{eqnarray}\label{partition}
{\cal M}=\displaystyle\sum_i p_i {\cal M}_i=\frac{1}{\displaystyle\sum_j e^{- {\cal S}_j}}\times\displaystyle\sum_i e^{- {\cal S}_i} {\cal M}_i,
\end{eqnarray} 
where ${\cal M}_i$ represents any model in space $\mathbb{H}$. If one can assign a Lagrangian ${\cal L}_i$ to each model ${\cal M}_i$, then the ensemble averaged Lagrangian will be
\begin{eqnarray}\label{lagrangian-general}
{\cal L}=\frac{1}{\displaystyle\sum_j e^{- {\cal S}_j}}\times \displaystyle\sum_i{\cal L}_i e^{- {\cal S}_i} .
\end{eqnarray}
Note that the action ${\cal S}_i$ is given by the Lagrangian ${\cal L}_i$ itself such that one can formally write ${\cal S}_i=V \times {\cal L}_i$ with $V$ the volume of spacetime. In the statistical mechanics formulation the partition function $e^{-\beta E_i}$ gives the probability of each micro-state, where $E_i$ is the energy of each micro-state and $\beta$ is the inverse temperature of the system. This fact shows a similarity between the temperature of space of models $\mathbb{H}$ and the inverse volume of spacetime.

In what follows we study the \"ubermodeling idea in the context of gravitational interactions and focus on higher order gravity models. We show that in four-dimensional spacetime the ensemble averaged model reduces to a specific form of $f(R,G)$ with $R$ and $G$ the Ricci and Gauss-Bonnet terms, respectively. We study the behavior of this specific model and show that it admits stable self-accelerating solutions. A non-trivial result is that this model behaves differently in three low-, intermediate- and high-curvature regions in comparison to GR. Finally, we conclude with a few remarks about the idea and possible future applications and directions. 


{\it\"Ubermodeling in higher order gravity framework}:
In what follows we assume the space $\mathbb{M}$ to contain higher order gravity models. It is well known that in this case the healthy subspace of $\mathbb{M}$, i.e. $\mathbb{H}$, is the family of Lovelock terms \cite{Lovelock}
\begin{eqnarray}\nonumber
&&L_0=\Lambda, \hspace{1cm} L_1=R,\\&&L_2=R^2-4R^{\mu\nu}R_{\mu\nu}+R^{\mu\nu\alpha\beta}R_{\mu\nu\alpha\beta},\hspace{.1cm} ...~.
\end{eqnarray}
According to our proposal the zeroth order Lovelock term, $L_0$, cannot belong to $\mathbb{H}$ because it does not result in dynamical equations of motion for the metric $g_{\mu\nu}$, and consequently, it is not a model of gravity on its own. Therefore, the ensemble averaged Lagrangian can be written as
\begin{eqnarray}\label{lagrangian0}
{\cal L}= \sqrt{-g}\bigg(\displaystyle\sum_{i=1}^N L_i e^{-\beta L_i}\bigg)\bigg/\bigg(\displaystyle\sum_{j=1}^N e^{-\beta L_j}\bigg),
\end{eqnarray}
where $\beta$ is a constant parameter. In four dimensions we have $L_i=0$ for all $i>2$, hence the Lagrangian above reduces to
\begin{eqnarray}\label{lagrangian}
{\cal L}= \sqrt{-g}\bigg[ \frac{M^2 R\hspace{.1cm} e^{-\beta M^2 R}}{e^{-\beta M^2 R}+e^{-\beta \alpha G}}+  \frac{\alpha G \hspace{.1cm}e^{-\beta \alpha G}}{e^{-\beta M^2 R}+e^{-\beta \alpha G}} \bigg],
\end{eqnarray}
where $R$ and $G$ are the Ricci scalar and the Gauss-Bonnet term, i.e. the first and second Lovelock terms, respectively\footnote{Our choice of pure $R$ and $G$ may arise a question that in general two combinations like $R+a\, G + b\, \Lambda$ and $G+a'\,R+b'\,\Lambda $ can be assumed as our basis where $a$, $a'$, $b$ and $b'$ are constants. It is easy to check that these models are also satisfy the stability condition (\ref{ghost-free-condition}). But in this work we prefer to keep $R$ and $G$ as the basis and work with less free parameters.}. Note that since the Lovelock terms are defined up to a constant we have then introduced $M$ and $\alpha$ so as to have the most general Lagrangians. $M$ has mass dimension $[M]$ and $\alpha$ is dimensionless, while the mass dimension of $\beta$ is $[M^{-4}]$. Also note that the Gauss-Bonnet term is a topological term if it appears alone in the action, which is not the case for a general form of $f(R,G)$. It is useful to reformulate the Lagrangian as
\begin{eqnarray}\label{lagrangian-partition}
{\cal L}=-\frac{\partial }{\partial \beta} \ln{\cal Z} \hspace{.2cm} for \hspace{.2cm}{\cal Z}=e^{-\beta M^2 R}+e^{-\beta \alpha G},
\end{eqnarray}
where $\cal Z$ is exactly the partition function of our model space $\mathbb{H}$. This reminds us of the ensemble averaged energy of any thermodynamical system, given by $\langle E \rangle=-\frac{\partial }{\partial \beta} \ln{\cal Z}$ with $\cal Z$ the partition function of the system. Hence, the ensemble averaged Lagrangian can be interpreted as the ensemble averaged energy of the model space $\mathbb{H}$.

The $f(R,G)$ models have been studied extensively in the literature \cite{fRG,Lovelock,deFelice-constraints}. In what follows
we study the properties of our model (\ref{lagrangian}) from different aspects. We will show that our model satisfies stability conditions and can have a self-accelerating solution. In addition, we show that the model has very specific predictions, making it distinguishable from $\Lambda$CDM, as well as other models of modified gravity. 

{\it Stability analysis}:
$f(R,G)$ models allow propagation of scalar, vector, and tensor modes, and their stabilities have been studied in the literature \cite{deFelice-constraints}. It has been shown that the stability of any $f(R,G)$ model depends quantitatively on satisfying the condition
\begin{eqnarray}\label{ghost-free-condition}
\Delta=f_{,RR} f_{,GG} - f_{,RG}^2=0,
\end{eqnarray}
which is a partial differential equation for $f(R,G)$. There are two trivial solutions for this equation, $R+ f(G)$ and $f(R)+G$. Interestingly, our model (\ref{lagrangian}) satisfies the above equation and does not belong to the family of trivial solutions. The above condition is necessary but not sufficient, and a model of $f(R,G)$ should satisfy other conditions in order to have healthy propagating modes. As it is shown in Ref.~\cite{deFelice-constraints}, in order to study other stability conditions one needs to specify the background. For our purpose we assume that our model can provide a Friedmann-Lema\^{i}tre-Robertson-Walker (FLRW) background solution, and in the next subsection we will show that this assumption is obtainable in our model. In order to utilize the formalism in Ref.~\cite{deFelice-constraints} it is better to define two new functions
$F\equiv\frac{\partial f}{\partial R}$ and $\xi\equiv\frac{\partial f}{\partial G}$. For a general form of $f(R,G)$, there are two propagating scalar modes with the speed of sound, in case of $\Delta=0$,
\begin{eqnarray}\label{cs-scalar}\nonumber
c_1^2&\approx&-\frac{1}{3}\bigg(16\dot{\xi}H^2\ddot{\xi}-64H^3\dot{\xi}^2-64\dot{\xi}^2\dot{H}H-12\dot{\xi}H^2F\\\nonumber&&\hspace{.8cm}-16\dot{F}H\dot{\xi}-16\dot{\xi}\dot{H}F+4\ddot{\xi}
	\dot{F}-3F\dot{F}\bigg)\bigg/\\&&\hspace{.8cm}\bigg(16H^3\dot{\xi}^2+4\dot{\xi}H^2F+4\dot{F}H\dot{\xi}+F\dot{F}\bigg),\\
c_2^2&\approx&\frac{\dot{p}}{\dot{\rho}},
\end{eqnarray} 
where $p$ and $\rho$ represent a perfect fluid. We recall that for the FLRW metric we have $R=6(2H^2+\dot{H})$
and $G=24 H^2 (H^2+\dot{H})$, where $H$ is the Hubble rate. The stability of $c_2^2$ does not depend on our $f(R,G)$ model but it depends on the matter field. It is easy to show that for the other mode, the speed of sound $c^2_1 \rightarrow 1$ for an approximate de Sitter solution. This shows that the scalar sector of our model is healthy for the interesting cosmological backgrounds. The stability of vector modes imposes another constraint, $F+4 H \dot{\xi}\geq 0$. It can be shown that for the regime of interest (i.e. an approximate de Sitter solution) $F+4 H \dot{\xi}\rightarrow\frac{\left(4+3 \sqrt[4]{e}\right) M^2}{4 \left(1+\sqrt[4]{e}\right)^2}$, which obviously satisfies the stability condition. As pointed out in Ref.~\cite{deFelice-constraints}, the stability condition of the vector modes imposes the absence of ghost in the tensor modes. However, one should check the speed of propagation of tensor modes which is given by $c_{\otimes}^2=c_{\oplus}^2=\frac{F+4\ddot{\xi}}{F+4 H \dot{\xi}}$, and for the regime of interest in our model we have $c_{\otimes}^2=c_{\oplus}^2 \rightarrow 1$, which means that the tensor modes propagate at the speed of light. This shows that for an approximate de Sitter background the perturbations are stable and our model is therefore viable.


{\it Cosmology}:
We have so far shown that perturbations are stable on an approximate de Sitter background. In what follows we show that such a background solution indeed exists in our model. The equations of motion for a general $f(R,G)$ can be read as (see Appendix A in Ref.~\cite{Lovelock})
\begin{eqnarray}\label{solutions-cos}\nonumber
&&\frac{\partial f}{\partial R}R_{\mu\nu}-\frac{1}{2}\bigg(f-G \frac{\partial f}{\partial G}\bigg)g_{\mu\nu}
+ \bigg(g_{\mu\nu} \Box -\nabla_\mu \nabla_\nu \bigg)\frac{\partial f}{\partial R}\\\nonumber
&&-4\bigg(G_{\mu\nu} \Box+\frac{1}{2}R \nabla_\mu \nabla_\nu - 2 R_{\alpha(\mu}\nabla_{\nu)}\nabla^\alpha
+\\&&\hspace{.8cm}(g_{\mu\nu}R_{\alpha\beta}+R_{\mu\alpha\beta\nu})\nabla^\alpha \nabla^\beta\bigg)\frac{\partial f}{\partial G}=0.
\end{eqnarray}
Finding exact solutions to the above equations for our model (\ref{lagrangian}) is not easy. Here, our goal is to illustrate that a suitable approximate solution can be found by employing symmetries as well as approximation methods; we leave the investigation of exact solutions to future work. First of all we simplify our Lagrangian by imposing the maximal symmetry,
\begin{eqnarray}
R_{\mu\nu\alpha\beta}=\frac{R}{D(D-1)}(g_{\mu\alpha}g_{\nu\beta}-g_{\mu\beta}g_{\nu\alpha}),
\end{eqnarray}
which is a symmetry of the FLRW metric; $D=4$ in our case. By looking at the above relation one can easily deduce that $G=\frac{1}{6}R^2$ and, consequently, the Lagrangian (\ref{lagrangian}) reduces to
\begin{eqnarray}\label{lagrangian-maximally-symmetric}\nonumber
{\cal L}= \sqrt{-g}\bigg[M^2 \frac{R\,e^{-\beta M^2 R}}{e^{-\beta M^2 R}+e^{-\beta \alpha \frac{R^2}{6}}}+ \frac{\alpha}{6} \frac{R^2\,e^{-\beta\alpha \frac{R^2}{6}}}{e^{-\beta M^2 R}+e^{-\beta\alpha \frac{R^2}{6}}} \bigg].
\end{eqnarray}
Since we are interested in the late-time acceleration, we can assume $R\ll 1$, which results in
\begin{eqnarray}
{\cal L}= \sqrt{-g}\bigg[\frac{M^2 R}{2}+R^2 \left(\frac{\alpha}{12}-\frac{\beta M^4}{4}\right)\bigg].
\end{eqnarray}
By fixing the lapse ($N=1$) one can obtain an equation of motion for the scale factor $a(t)$ as
\begin{eqnarray}\nonumber
&&3 \left(\alpha-3 \beta  M^4\right) a'^4+12 a \left(3 \beta  M^4-\alpha\right)
	a'^2 a''-M^2 a^2 a'^2\\\nonumber&&+a^2 \left(3 \left(3 \beta  M^4-\alpha\right) a''^2+4
	a^{(3)} \left(3 \beta  M^4-\alpha\right) a'\right)=0,
\end{eqnarray}
where $'$ denotes a derivative with respect to time. In order to check whether a de Sitter metric can be a solution of the above equation we plug $a(t)=e^{\lambda t}$ into the equation and reach
\begin{eqnarray}
\bigg(16 (3 \beta   M^4-\alpha) \lambda ^2- M^2\bigg)\lambda ^2=0,
\end{eqnarray}
which has a non-trivial solution $\lambda=\pm\frac{M}{4 \sqrt{3 \beta  M^4-\alpha}}$. We choose the positive branch
for the late-time acceleration, and we need to impose $3 \beta M^4>\alpha$ for a valid solution. For a very specific range of parameters one can assume $\beta M^4\gg\alpha$ and, consequently, $\beta\sim (\lambda M)^{-2}$. Now by assuming $M=M_{\text{Pl}}$ and $\lambda=H_0$, with $M_{\text{Pl}}$ and $H_0$ the Planck mass and present Hubble rate, respectively, we have  $\beta \approx M_{\text{EW}}^{-4}$, where $M_{\text{EW}}$ is the electroweak scale. Note that we interpreted $\beta$ as a volume of spacetime in the introduction, which is $H_0^{-4}$ here. One can assume $\beta\sim H_0^{-4}$, and therefore $M\sim H_0$, implying $\lambda\sim H_0$. The price to pay is a fine-tuned coupling between the matter field and gravity since $M$ cannot play the role of $M_{\text{Pl}}$, i.e. the true coupling constant. 

\begin{figure}
	\centering
	\includegraphics[width=.9\linewidth]{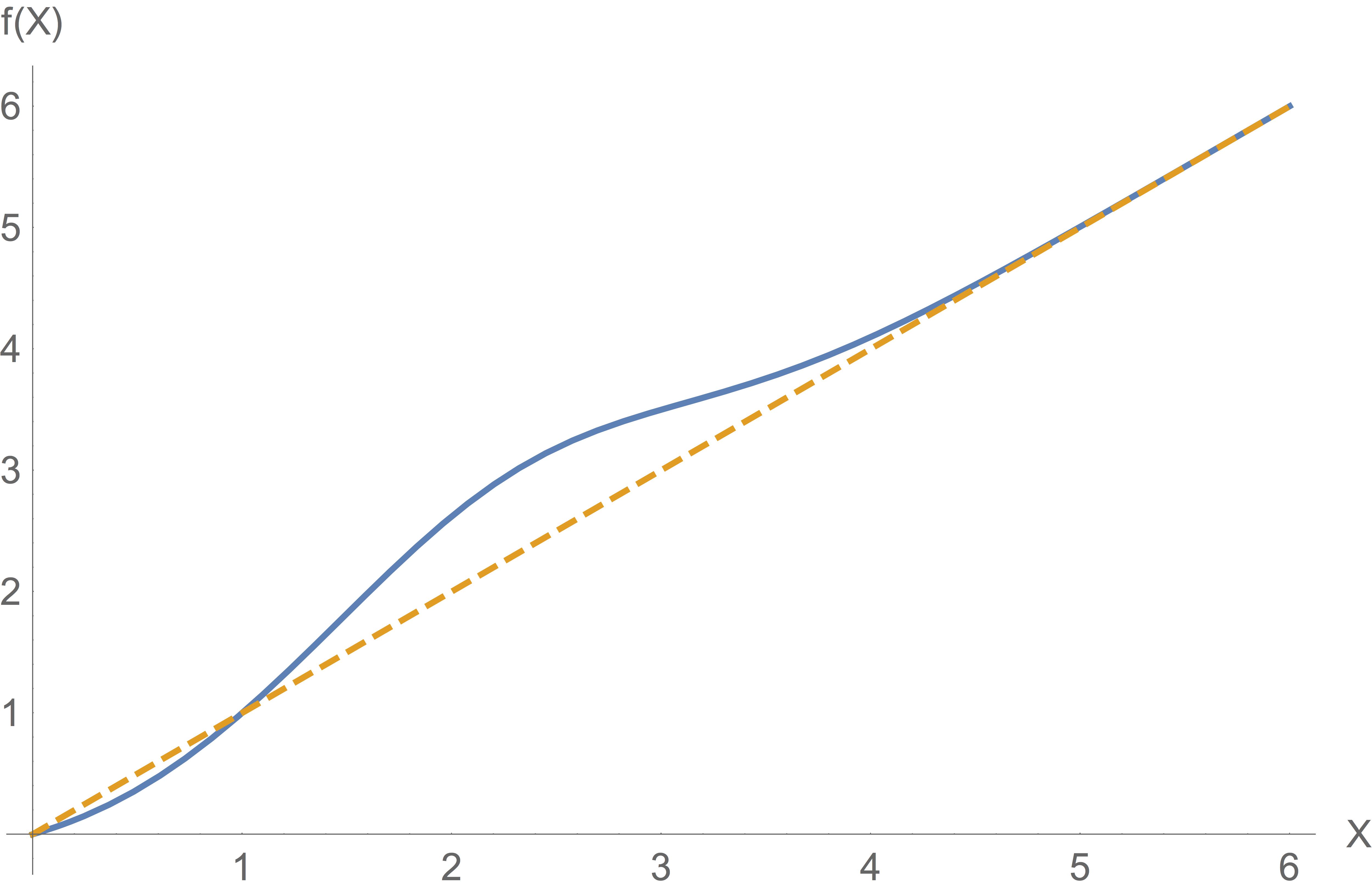}
		\caption{In this plot the solid line represents $f(X)$ for $b=0.4$ and $\alpha=1$, and the dashed line corresponds to GR. For high-curvature regions,
		both models behave identically. Our modified-gravity model however predicts a weaker gravitaionl force for low-curvature regions which can be interpreted as a repulsive fifth force responsible for dark energy. The smoking gun of the model is the stronger gravitational force in an intermediate-curvature region which affects the growth rate of structure formation.}
	\label{fig:RG}
\end{figure}

\begin{figure}
	\centering
	\includegraphics[width=0.9\linewidth]{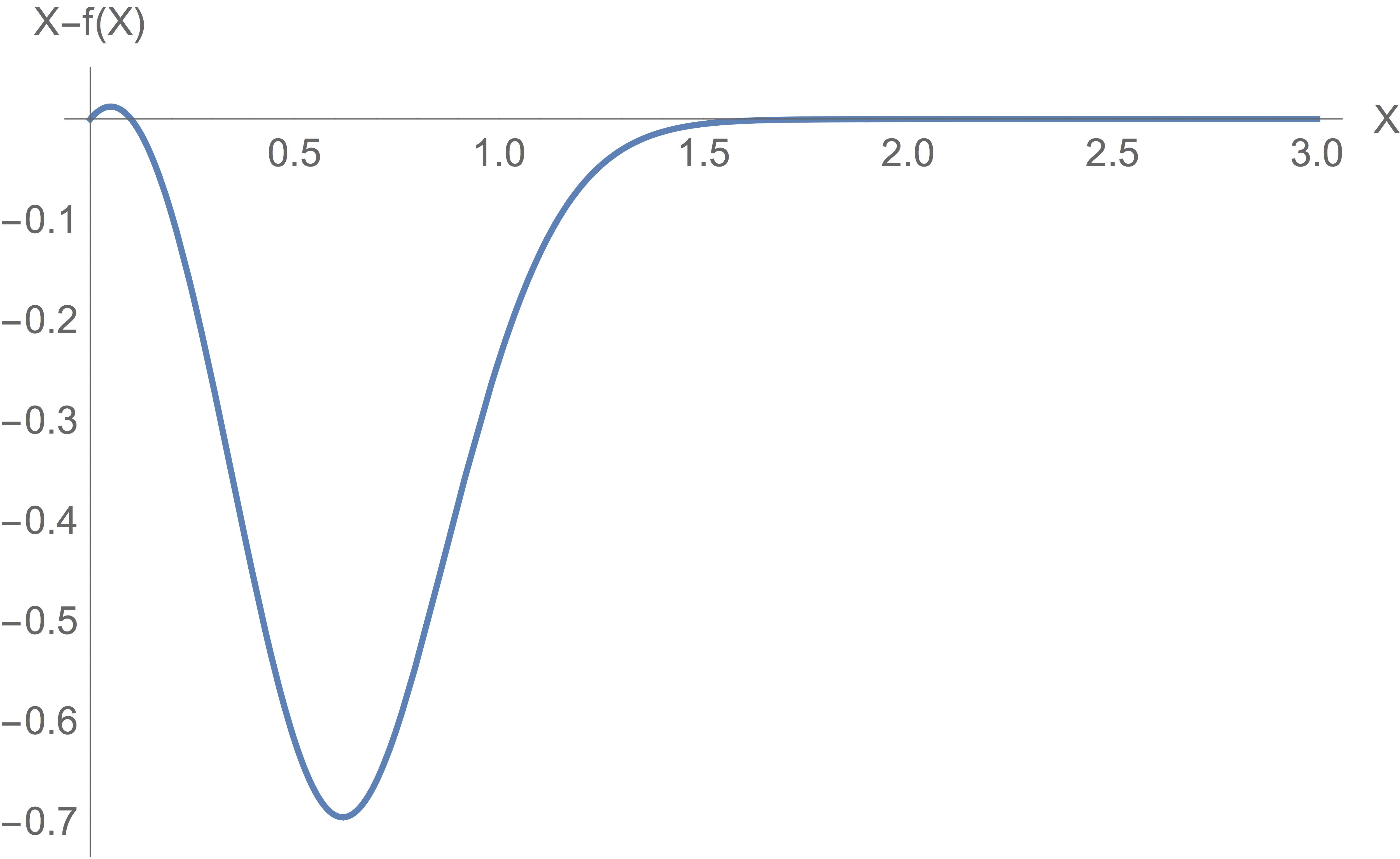}
	\caption{This figure shows the difference between GR and our model, $X-f(X)$, for $b=0.4$ and $\alpha=10$. Increasing $\alpha$ makes low- and intermediate-curvature regions smaller and shifted towards the lower curvature regimes.}
	\label{fig:RG-EH}
\end{figure}

{\it General properties}:
Let us now show that our model has remarkable predictions which make it distinguishable from $\Lambda$CDM and other theories of modified gravity. To do so, we focus on general properties of our model by assuming $R\sim X M^{2}$ and $G\sim X^2 M^{4}$, where $X$ is a representative of the curvature normalized by the scale $M$. The Lagrangian now takes the form
\begin{eqnarray}\label{fX}
	\frac{1}{M^4(-g)^{\frac{1}{2}}} {\cal L}= \frac{X e^{-b X}+\alpha X^2 e^{-b \alpha X^2}}{ e^{-b X}+e^{-b \alpha X^2}}=f(X),
\end{eqnarray}
where we have defined the dimensionless quantity $b\equiv\beta M^4$. Fig.~\ref{fig:RG} shows our model $f(X)$ versus GR, in solid and dashed lines, respectively. It is obvious that the behavior of our model in comparison to that of GR depends on the curvature. In the high-curvature regime, i.e. $X\gg 1$, one cannot distinguish the model from GR. In fact, this means that our model successfully satisfies the very strong constraints from local tests of gravity. In the opposite regime, i.e. the low-curvature region $X\ll 1$, the gravitational force is weaker than in GR. This can be interpreted as a repulsive fifth force which is responsible for the late-time acceleration of the Universe. However, as it is clear from the plot, the strength of the fifth force depends on the curvature, which can be seen as a fingerprint of our model. But the smoking gun is in the intermediate-curvature regime where, the gravitational force in our model is stronger than in GR. It seems that our model has a kind of chameleonic behavior such that the strength of gravity is sensitive to the curvature. The properties of our model have observational implications, for example for the growth rate. Specifically, the stronger gravity in the intermediate regime speeds up the collapse process of dark matter at higher redshifts. This means that one should expect more massive objects at higher redshifts in comparison with what is predicted by $\Lambda$CDM. This feature can be observed with the next generation of galaxy surveys, such as the {\it SKA} and {\it Euclid}, which will probe higher redshifts with high resolution.  In order to quantify the properties of our model, it is more suitable to work with $X-f(X)$, which is plotted in Fig.~\ref{fig:RG-EH}. It is easy to check that $X-f(X)$ has two solutions, one at $X=0$ and the other one at $X_b=\frac{1}{\alpha}$. The $X_b=\frac{1}{\alpha}$ solution is at the border of low- and intermediate-curvature regions.
The scale of the intermediate-curvature region is at the local minimum of $X-f(X)$. In order to find this point, we can look at the first derivative of $X-f(X)$,
\begin{eqnarray}\nonumber
\frac{d }{d X} \bigg(X-f(X)\bigg)\propto&&\\&&\hspace{-3cm} \left(1-2 \alpha  X\right) \left(e^{\alpha  b X^2} \left[b X \left(1-\alpha 
X\right)+1\right]+e^{b  X}\right)
\end{eqnarray}
which has an obvious solution at $X_{\text{max}}=\frac{1}{2\alpha}$ located in the low-curvature region. Hence, the local maximum of $X-f(X)$ is exactly at the half of the the location of the border between low- and intermediate-curvature regions, $X_b$. The other solution occurs at $X_{\text{min}}=\frac{1\pm \sqrt{1+\frac{4 \alpha  c}{b}}}{2 \alpha }$, which is the scale of the intermediate-curvature region, where $c=1+W(1/e)$, $W$ is the Lambert $W$ function, and $W\left(\frac{1}{e}\right)\approx 0.278465$. It is clear that increasing $\alpha$ decreases the value of $X_{\text{min}}$, as well as $X_b$, as one can see by comparing both Figs.~\ref{fig:RG} and~\ref{fig:RG-EH}.
We previously showed that the observable cosmological constant $\frac{\lambda^2}{M^2}\sim \frac{1}{b}$ needs $b\gg \alpha$, which means that the cosmological constant should happen in the low-curvature region, i.e., $\frac{\lambda^2}{M^2}\ll \frac{1}{\alpha}$, which is fully compatible with the analysis of $X-f(X)$. Note that for the observed cosmological constant one needs to go very deeply into the low-curvature regime, which means very large values for $\alpha$. This implies that the low- and intermediate-curvature regimes happen for very small $X\ll M$. Now by tuning $M=M_{\text{Pl}}$ there is enough room for $X$ to pass the local gravity tests before reaching the quantum gravity regime.

{\it Conclusions and outlook}:
The \"ubermodeling idea introduced in this paper gives us an opportunity to look from a different angle at the theoretical model-building procedure. Every theoretically consistent model plays a role in the construction of the actual model for describing nature, and is weighted by a partition function inspired by the Euclidean path-integral formalism. In the higher order gravity framework, this idea suggests a reason why GR is the dominant model for high-curvature regimes while the Gauss-Bonnet term cannot be ignored for low-curvature regimes. For the deduced $f(R,G)$ model (\ref{lagrangian}) we need to study carefully the exact background solutions as well as the behavior of perturbations. The smoking gun of the model proposed here is in the intermediate-curvature region, where gravity is stronger than what GR predicts. As far as we know, this is a unique property of our model that can be interpreted (at least partially) as an explanation for dark matter. This feature, which should be studied carefully, can affect the growth rate of cosmic structure.

A question about our model is its UV-completion i.e. how the Lagrangian (\ref{lagrangian0}) or (\ref{lagrangian}) should be quantized as well as how it couples to (quantum) matter. The proposed effective action, once coupled to matter fields, needs
quantization. One can implement this, e.g., in the path integral
formulation. This however, would likely introduce new divergences
beyond and in addition to the once existing in the Euclidean path
integral formulation of GR, owing to the fact that the
Euclidean actions for gravity (including GR action) are not positive
semi-definite. Whether these new divergences make our proposal less
manageable than the Euclidean path integral formulation of GR remains
to be seen.

An interesting direction of research is to apply the \"ubermodeling idea to other theoretical frameworks, such as massless/massive gravity~\cite{massive} and multi-metric models~\cite{multimetric}. Another interesting idea is to show why Riemannian geometry is favored by nature amongst all the possible Weyl geometries \cite{eps}. This idea can also be employed in constructing models other than for gravity, e.g. gauge field theories. A question that has been discussed in the literature is why the physics of nature is explained by gauge symmetries. In the context of \"ubermodeling idea one can consider all the models with and without gauge symmetry and try to see if \"ubermodeling automatically shows that gauge invariance is more probable.  In Ref.~\cite{nielsen}, it has been shown that a model with gauge invariance is emergent even if there is gauge non-invariance at small scales. Perhaps the \"ubermodeling idea is rich enough to answer why nature is natural?

\vskip 0.05in
{\it Acknowledgments}:
I would like to thank Y. Akrami, S. Baghram, F. K\"onnig, I. Sawicki and M. M. Sheikh-Jabbari for very fruitful and instructive discussions, as well as their comments on the draft. I also thank A. De Felice, R. Durrer, M. Kunz, M. Memarzadeh and S. Rouhani for their comments and discussions. I am also grateful to the University of Geneva, School of Astronomy and School of Physics at IPM for their hospitality during the completion of this work. I would like to thank anonymous referee for her/his useful comments.








\end{document}